\documentclass[journal]{vgtc}
\ifpdf
  \pdfoutput=1\relax                   
  \usepackage{graphicx}                
  \DeclareGraphicsExtensions{.pdf,.png,.jpg,.jpeg} 
\else
  \usepackage{graphicx}                
  \DeclareGraphicsExtensions{.eps}     
\fi%

\usepackage{microtype}                 
\PassOptionsToPackage{warn}{textcomp}  
\usepackage{textcomp}                  
\usepackage{mathptmx}                  
\usepackage{times}                     
\usepackage{cite}                      

\usepackage{amsmath}
\usepackage[ruled,vlined,linesnumbered]{algorithm2e}

\makeatletter
\newcommand{\removelatexerror}{\let\@latex@error\@gobble}
\makeatother

\usepackage{soul}
\usepackage{color}

\vgtccategory{Research}

\vgtcinsertpkg

\usepackage{multirow}
\usepackage[table]{xcolor}

\newcommand{\Cpp}{C\raise .2ex \hbox{++}}

\title{Further Towards Unambiguous Edge Bundling: Investigating Power-Confluent Drawings for Network Visualization}
\author{Jonathan X.\ Zheng, Samraat Pawar, and Dan F.\ M.\ Goodman}
\authorfooter{
\item
J.\ X. Zheng is with Imperial College London. Email: jxz12@ic.ac.uk
\item
S.\ Pawar is with Imperial College London.
\item
D.\ F.\ M.\ Goodman is with Imperial College London.
}

\abstract{
Bach et al.~\cite{bach} recently presented an algorithm for constructing confluent drawings, by leveraging power graph decomposition to generate an auxiliary routing graph. We identify two issues with their method which we call the node split and short-circuit problems, and solve both by modifying the routing graph to retain the hierarchical structure of power groups. We also classify the exact type of confluent drawings that the algorithm can produce as `power-confluent', and prove that it is a subclass of the previously studied `strict confluent' drawing. A description and source code of our implementation is also provided, which additionally includes an improved method for power graph construction.}
\teaser{
    \centering
    \includegraphics[width=.95\textwidth]{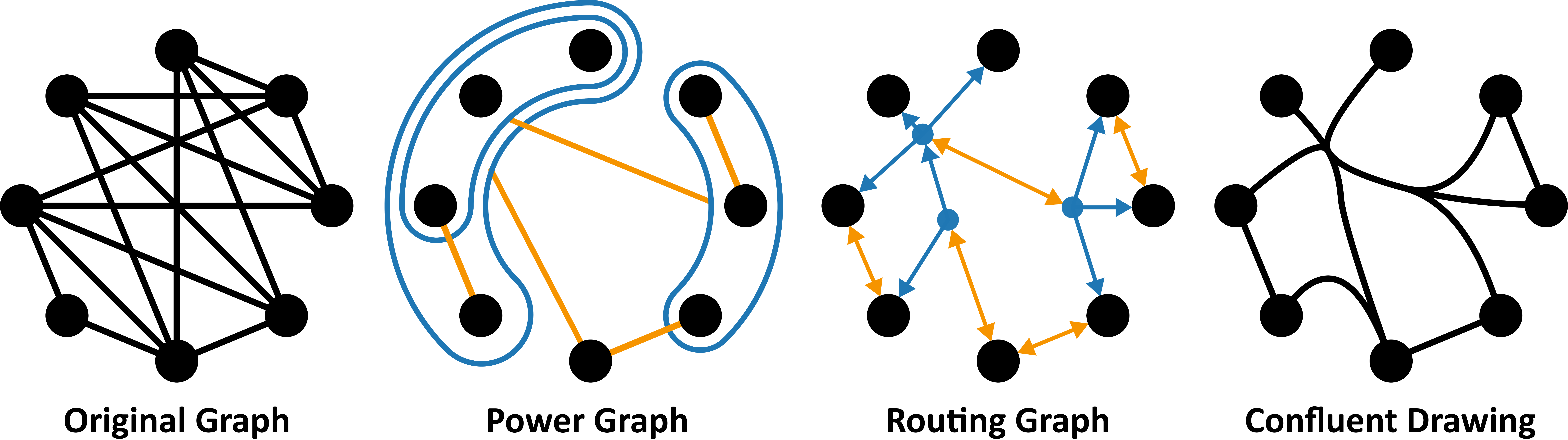}
    \caption{An illustration of our solution to the issues with the algorithm of Bach et al.~\cite{bach}, using their original example graph (Fig.~2 in~\cite{bach}).
    From left to right: a conventional node-link diagram with vertices arranged around a circle; its power graph decomposition, where edges are compressed by grouping similar nodes together; our novel equivalent routing graph, where the nested structure of groups is represented by a directed tree; the resulting confluent drawing, where original edges are threaded through the routing graph to form bundled junctions.
    }
    \label{teaser}
}

\begin{document}

\firstsection{Introduction}
\maketitle
Confluent drawing is a graph drawing technique that eliminates crossings by allowing edges to overlap, as long as the join is smooth.
This is analogous to junctions on a train track that allow carriages to switch directions without stopping.
The original definition (Dickerson et al.~\cite{dickerson}) of a confluent drawing~$A$ for a graph~$G$ is:
\begin{itemize}
    \item There is a one-to-one mapping between the vertices in $G$ and $A$, so that, for each vertex $v\in V(G)$, there is a corresponding vertex $v'\in A$, which has a unique point placement in the plane.
    \item There is an edge $(v_i,v_j) \in E(G)$ iff there is a locally-monotone curve (defined as having no self intersections or sharp turns) $e'$, connecting $v_i'$ and $v_j'$ in $A$. 
    \item $A$ is planar. That is, while locally-monotone curves in $A$ can share overlapping portions, no two can cross.
\end{itemize}
Bach et al.~\cite{bach} propose to construct such a drawing by converting a power graph decomposition into an auxiliary routing graph. Then, for each adjacency, the graph-theoretic shortest path through the auxiliary graph is used as the sequence of control points for a B-spline~\cite{sederberg}. This process is illustrated in Fig.~\ref{teaser}; see Bach et al.~\cite{bach} for detailed background and definitions.

We find that the combination of splines and shortest paths introduces problems not fully explored by the original authors, causing the resulting drawings to violate the second and third conditions in the above definition. We solve these problems, and identify the subclass of confluent drawings the algorithm can produce as \emph{power-confluent}. Our solution guarantees that the second condition is always satisfied, but still cannot guarantee the third, as discussed in Section~\ref{classification}.
Source and pseudocode for our implementation is provided in Section~\ref{implementation}.

\subsection{B-splines}
\label{bsplines}
We first identify issues regarding the use of B-splines (of degree $p=3$ in the source code of~\cite{bach}, although it is not specified in the paper) for interpolating control points.
These were likely chosen because they satisfy the convex hull property (which prevents crossings at shared control points; see Jia et al.~\cite{jia} for an example of poorly implemented B-splines that do not satisfy this property). They also offer local control (i.e.\ moving a control point only affects the surrounding $p+1$ segments), which guarantees that splines that share enough intermediate control points will overlap.
Local control is what makes it possible for drawings to be confluent; with the right routing graph, it is possible for edges to share enough control points so that they produce the overlapping portions that are required for a confluent drawing. Specifically, for two curves to be guaranteed to overlap they must share $p$ or more control points (see Section~\ref{splicifics} for a proof with a more detailed explanation).

\begin{figure}
    \centering
    \includegraphics[width=\linewidth]{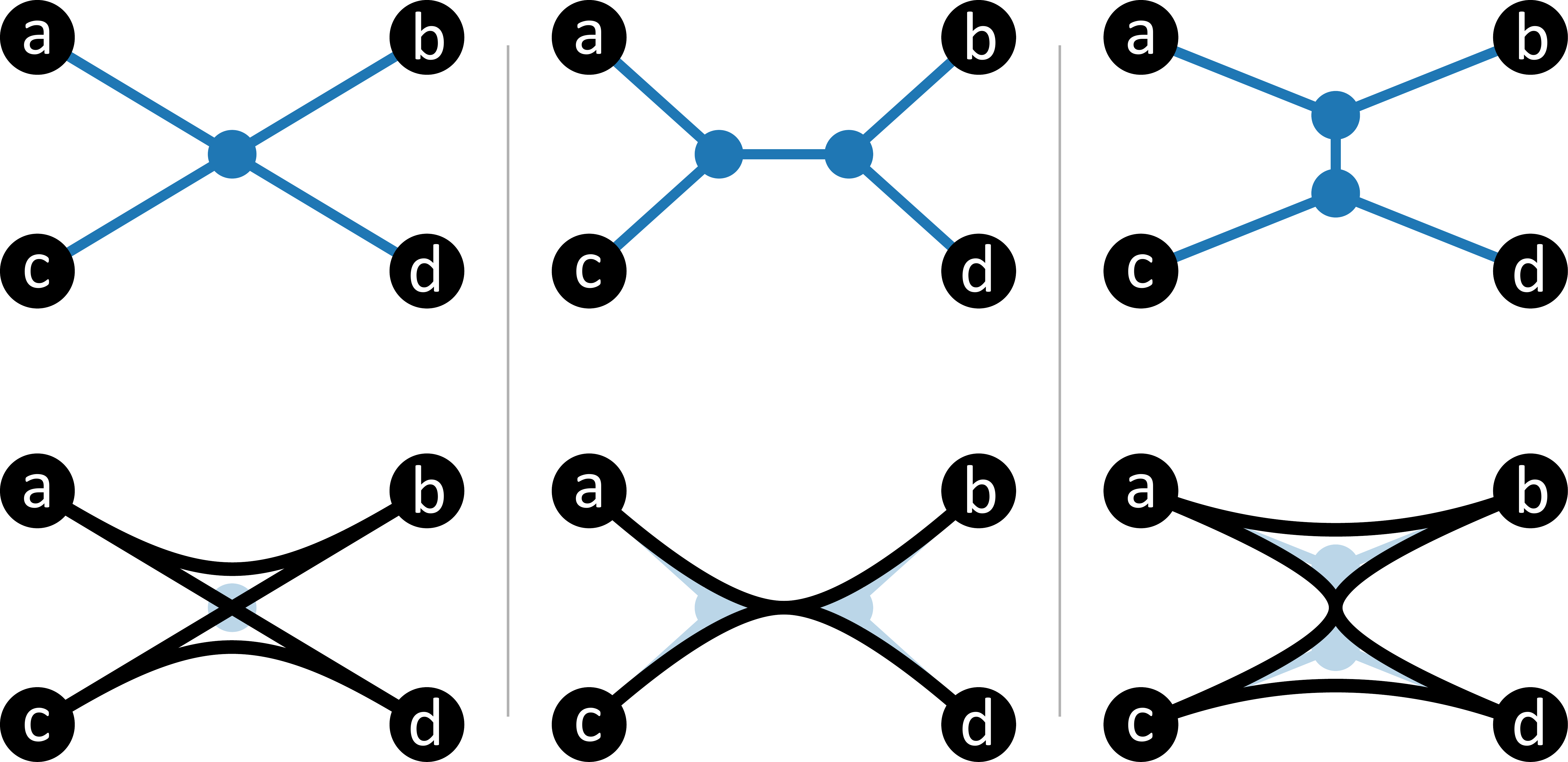}
    \caption{Examples to show how the direction of node splitting can affect the graph $K_{2,2}$. The first column shows a standard situation where a split is required to make edges overlap, the second shows the correct split, and the third shows that if the node is split the other way, then the edges $\{a,c\}$ and $\{b,d\}$ are falsely introduced.
    }
    \label{nodesplit}
\end{figure}

There are two problems with using B-splines in this context. The first is that splines that share fewer than $p$ control points will \emph{not} overlap, but sharing even a single routing node should indicate a bundled junction.
The authors recognized this, calling it the `crossing artifact'~\cite[Fig.~4]{bach}, and fixed it by splitting routing nodes into two: one for incoming and one for outgoing edges.
The intuition behind this splitting is correct, as it introduces another shared control point that tightens the bundle, merging the crossed edges into a junction. However, their exact description contains an ambiguity in the context of undirected graphs, as it is not specified how to identify edges as incoming or outgoing. This is problematic as an incorrect split may introduce errors into the resulting drawing, as illustrated in Fig.~\ref{nodesplit}.
We resolve this ambiguity, and the following related problem, in Section~\ref{solution}.

The second problem is also caused by local control but has an opposite result: that splines will \emph{always} overlap if they share $p$ or more control points.
For some routing graphs, such as the examples shown in Fig.~\ref{overlap}, splines may overlap so as to create the visual impression that extra edges, not present in the original graph, exist. This violates the second condition in the confluent definition.
We find that such a routing graph can result from the power-to-routing graph conversion, as explained in Section~\ref{shortcircuits}.

\subsubsection{Spline specifics}
\label{splicifics}
This section explains the construction of B-splines in more detail, and may be skipped if the reader is not interested in the more technical details.
A good introduction to B-splines and their decomposition into segments can be found in Sederberg~\cite{sederberg}.

\begin{figure}
    \centering
    \includegraphics[width=\linewidth]{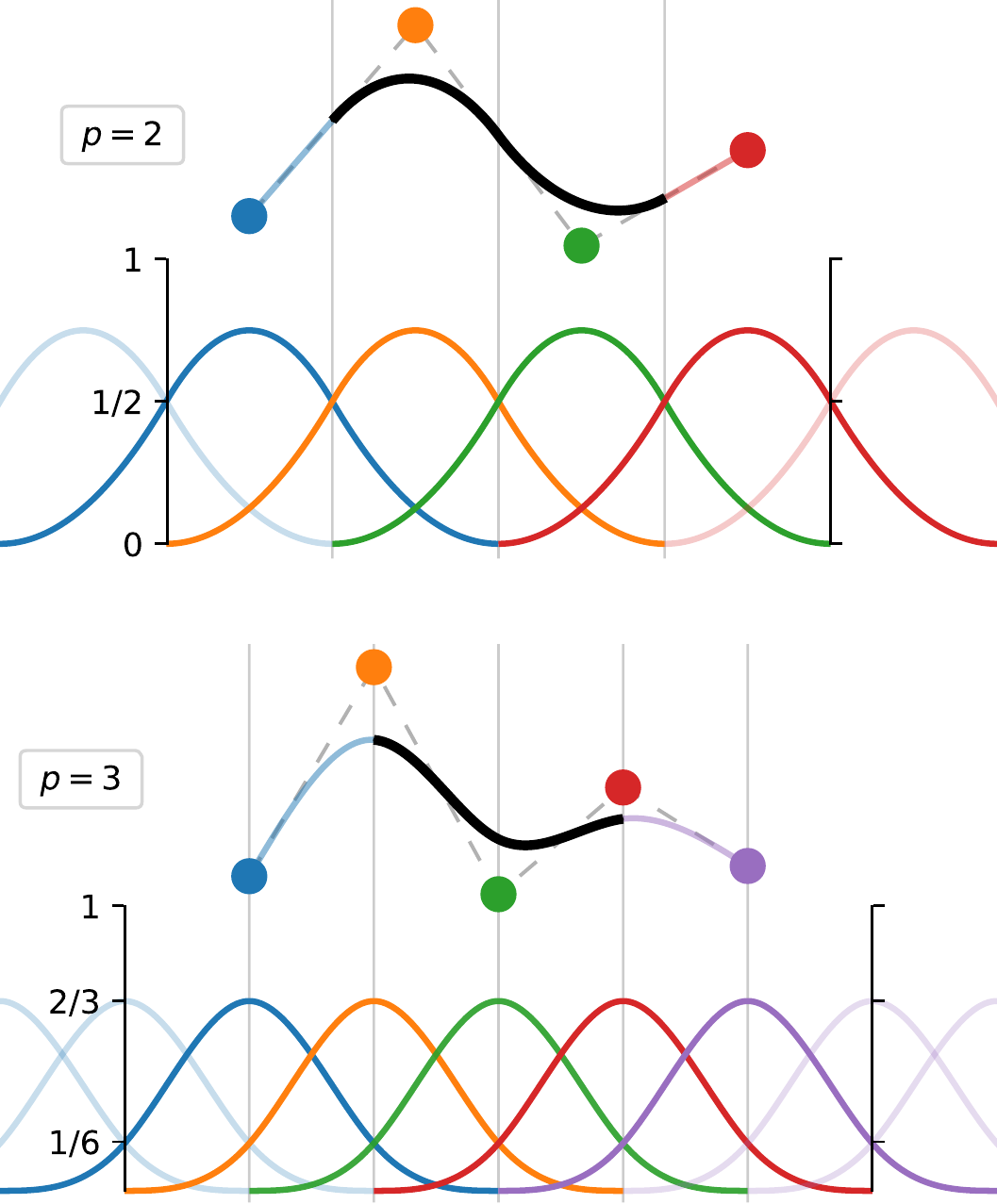}
    \caption{Examples of two open uniform B-splines and their basis functions below. Control points are spaced evenly along the x-axis such that knots (vertical lines) align with their associated segments. The top spline is quadratic (degree $p=2$), while the bottom is cubic ($p=3$). In both, the left- and rightmost basis functions go to zero at the middle knot, and so another curve that shares the middle $p$ control points is guaranteed to overlap exactly at that knot. Note that the sum of basis functions must always add up to one, so to attach a curve to its endpoints, the final control point is repeated $p$ times. The curves there are also slightly transparent, to match the corresponding repeated basis functions.}
    \label{basis}
\end{figure}

We first prove that splines of degree $p$ must share at least $p$ control points to overlap. A B-spline is a parametric curve that interpolates a number of control points by summing up contributions from a basis function for each (Fig.~\ref{basis}). As a consequence, the spline is constructed as a series of polynomial
curves, known as segments, which each satisfy
the local control property, meaning that each segment is affected by the basis functions of only the surrounding $p+1$ control points. These basis functions are continuous, so at the exact point at which two segments join (known as a knot) the curve cannot simultaneously be affected by the two furthest control points affecting the segments on either side. The remaining control points at the knot are therefore the last $p$ control points of the left segment, which overlap with the first $p$ control points of the right segment.
This proof is illustrated in Fig.~\ref{basis}. As a result, splines used must be quadratic ($p=2$) for node splitting (Fig.~\ref{nodesplit}) to work as intended.

The values of knots determine the parametric intervals over which the segments span, and the above proof assumes a `uniform' knot vector, where uniform is defined as having knots spaced along even parametric intervals. This ensures that splines with overlapping control points also have overlapping knots.
To additionally ensure that splines are connected to the first and last control point, the knot vector must also be `open', i.e.\ it contains $p$ repeated knots at each end.
However, there are two ways of doing this: the first retains the same total number of knots, for example a quadratic spline ($p=2$) with 4 control points and the standard knot vector $\langle0,1,2,3,4\rangle$ becomes $\langle0,0,1,2,2\rangle$, which eventually reduces to a B\'ezier curve~\cite{sederberg}. The second appends $p-1$ knots at either end, where $\langle0,1,2,3,4\rangle$ becomes $\langle0,0,1,2,3,4,4\rangle$, effectively duplicating the first and last control points.
This second method is illustrated in Fig.~\ref{basis}.

In practice this second method makes the rendered curve hug its path through the routing graph closer than the first. In the case of cubic splines ($p=3$), this means that edges come sufficiently close to look bundled, even though the above proof demonstrates that the degree must be $p=2$ for curves to fully overlap (unless routing nodes are split twice to guarantee three shared control points).
This appears to be the choice of the original authors~\cite{bach} in their provided source code.
The benefit of cubic splines is that they are $C^2$ continuous, i.e.\  have no sudden jumps in curvature, and having this extra smoothness is more aesthetically pleasing. Regardless, for all figures here except for Fig.~\ref{basis} we use quadratic splines, and with the first method of joining to end points.

Note that drawing a spline for every edge is not the only way to render the graph, and also introduces a great deal of redundancy due to the large amount of overlap between edges. If a purely confluent drawing is desired, then it is not necessary to redraw overlapping segments.
Allowing bundles to be relaxed~\cite[Fig.~18]{bach} is, however, a useful option that rendering each edge does provide.

\subsection{Short-circuits}
\label{shortcircuits}

\begin{figure}
    \centering
    \includegraphics[width=\linewidth]{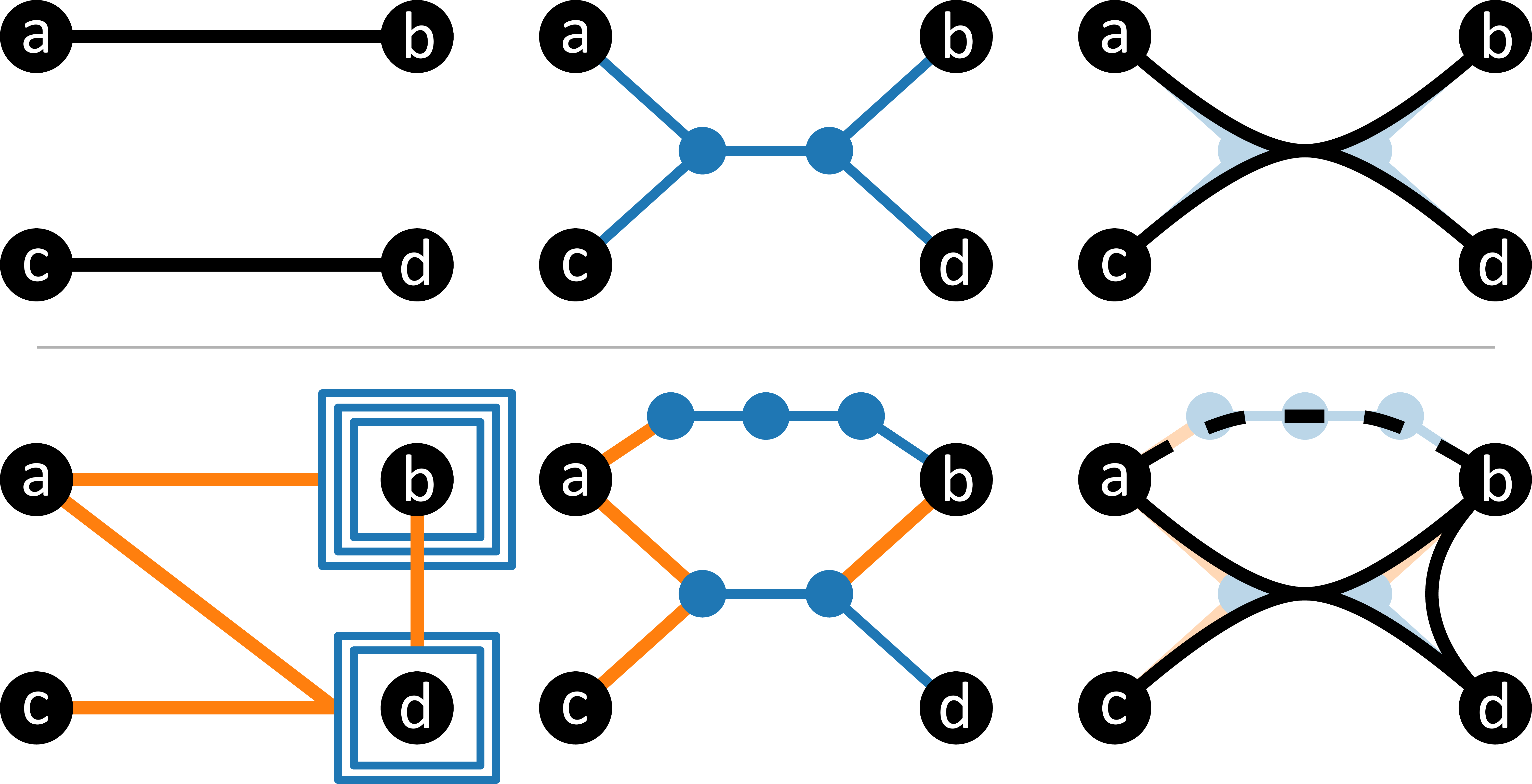}
    \caption{Examples of how absent edges can appear to exist.
    Top: a toy example of a hypothetical routing graph that causes edges $\{a,d\}$ and $\{c,b\}$ to both erroneously appear to exist.
    We use two intermediate routing nodes because that is the minimum number of shared control points required for quadratic splines to overlap (Section~\ref{splicifics}).
    Bottom: an example of a power graph decomposition that causes a similar ambiguity, where $c$ appears connected to $b$ because the edge $\{b,d\}$ causes a short-circuit through the nested structure of power groups (represented by blue boxes), causing $\{a,b\}$ to be routed through the wrong direction.
    The correct direction that would not cause an issue is shown as a dashed line along the top.
    Note that every power group in this example is intentionally redundant for illustrative purposes; a full version of this example can be seen in Fig.~\ref{overlap_sm}.
    }
    \label{overlap}
\end{figure}

Here we will explain why routing edges through their graph-theoretic shortest paths in the routing graph can introduce false adjacencies.
A power graph is an extension to the conventional node-link diagram, that compresses the number of edges by grouping similar vertices together into \emph{power groups}, and merging edges among group members that share the same target vertex into a single \emph{power edge} instead (see the leftmost two drawings in Fig.~\ref{teaser}).
This is then converted into a routing graph by (a) connecting the members of each power group to a routing node corresponding to the group, and (b) connecting pairs of routing nodes whose corresponding power groups are connected by power edges (see~\cite[\S~3.1]{bach} for more detail).
Since the original edges do not exist anymore in this auxiliary graph, they are instead drawn back on top by finding the graph-theoretic shortest path between the vertices on either end of the edge, using the nodes on this path as the control points for a B-spline.
However, because (a) and (b) both result in edges in the routing graph, with nothing to differentiate between them, this can cause a \emph{short-circuit} effect, that potentially introduces false adjacencies into the resulting drawing, as shown in Fig.~\ref{overlap}.

This effect can be explained as follows. The structure of groups within a power graph can be represented as a tree, where groups are represented by branches and vertices by leaves.
Trees are geodetic (i.e.\ there exists a unique shortest path between any pair of vertices), but the connections introduced by power edges can act like bridges between branches, to invalidate this and produce ambiguity either in the choice of path (if the shortest paths are equal) or in which edges exist at all, by routing splines in the wrong direction entirely.
The bottom row in Fig.~\ref{overlap} shows a simple example of how this can happen. While it may seem as if our counter-example is contrived and should not ever appear due to the redundant nested structure of power groups, a similar pattern arises from the optimal decomposition of a clique (see Fig.~\ref{overlap_sm}).
The key detail here is that only one power edge should ever be traversed for any given adjacency, which is guaranteed by our solution, described in the following section.

\section{Hierarchical routing}
\label{solution}
\begin{figure}
    \centering
    \includegraphics[width=\linewidth]{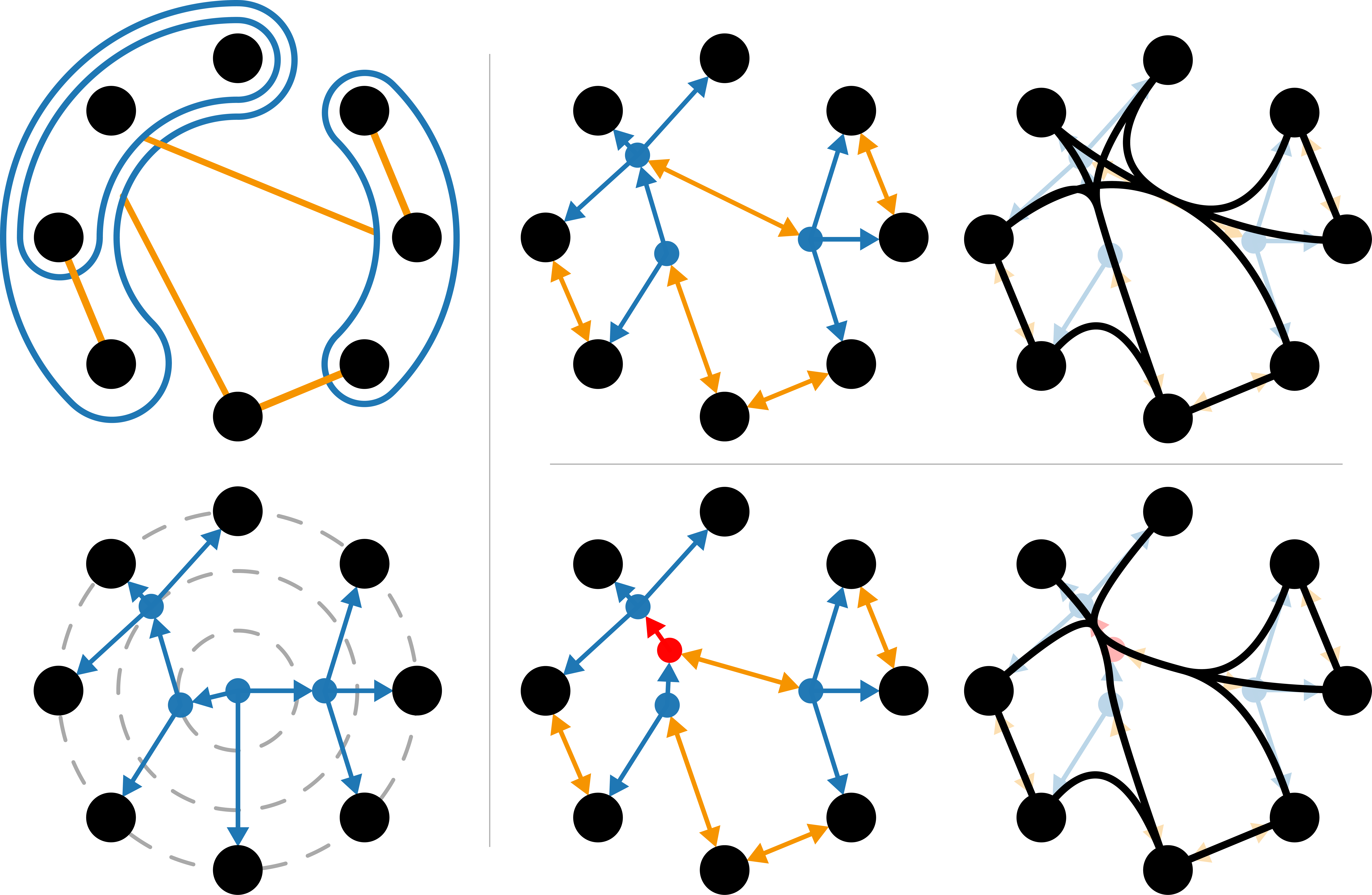}
    \caption{An illustration of the solution to both the node splitting and short-circuit ambiguities (Section~\ref{solution}). We use the same example graph as in~\cite[Fig.~2]{bach} and here Fig.~\ref{teaser}, shown in a radial layout to emphasize the tree structure of power groups.
    The leftmost graphs are a power graph (top), and a tree that represents the hierarchy of power groups (bottom), but without power edges.
    The middle column contains the resulting routing graphs without node splitting (top) and with (bottom, split indicated in red). Note that these do not include or require the root of the tree.
    The final column displays the resulting bundled layouts.
    }
    \label{radial}
\end{figure}

The solution to these problems is to retain the hierarchical structure of power groups as a directed tree, with all routing edges directed towards the leaves. We then add the power edges back, except as a special type of edge that is incoming at both ends (the purpose of which will soon be made clear). If we finish by discarding the root of the tree, we are left with the exact same routing graph as before, except now with all routing edges explicitly directed (Fig.~\ref{radial}, middle column).
Note that this also means all adjacency information is now preserved in the routing graph, such that the original graph can be recovered.

To draw the adjacency edges back on top, we can now forgo any shortest path calculations. Instead, for each power edge, we perform a depth-first search for all child leaf nodes starting from both ends of the power edge, and concatenate the path to each leaf from one end to the reversed path to each leaf from the other end. Every concatenated path is then used as the sequence of control points for a spline.
These paths are now guaranteed to be unique because the routing graph is effectively a tree, due to power edges being incoming at both ends to prevent their traversal. Since only the one correct power edge is traversed for each spline, the short-circuit problem described in Section~\ref{shortcircuits} is alleviated.

Imposing this directionality on the routing graph also fixes the node splitting ambiguity in Section~\ref{bsplines}, guaranteeing that the split occurs in the correct direction.
This works because every routing node is the boundary between two sides of a biclique, equivalent to a single bundled junction; the explicit direction of routing edges now encodes the orientation of the bundle itself.
This entire process is illustrated in Fig.~\ref{radial}, and pseudocode for the resulting algorithm can be seen in Fig.~\ref{pcd_pseudo}.

\subsection{Implementation}
\label{implementation}
To aid future work, here we complete the description for the algorithms used in the current paper\footnote{Source code is available at \url{www.github.com/jxz12/pconfluent}.}, and also provide pseudocode for the benefit of the reader. 
Note that the following description is specific to undirected graphs; we explain the changes required for directed graphs in Section~\ref{directed}.

\subsubsection{Improved power graph decomposition}
Here we outline our power graph generation algorithm, which includes one improvement from the one used by Bach et al.~\cite{bach}, originally developed by Dwyer et al.~\cite{beamsearch}. The algorithm works by also taking advantage of the fact that the hierarchy of power groups can be represented as a tree, with vertices as the leaves. It therefore first initializes every vertex as the sole member of a trivial power group---known as a module---and initializes a set of neighbours for each module based on the edges of the graph.
It then greedily merges pairs of modules at a time, picking the merge that eliminates the most edges at each step, until no more edges can be eliminated.
The number of eliminated edges for any given merge is given by
\begin{equation}
    \kappa_\cap(m, n) = |N(m)\cap N(n)|
\end{equation}
where $m$ and $n$ are modules and $N$ are their neighbour sets.
This is almost the same as $nedges(m,n)$ in~\cite{beamsearch}, except as a positive score rather than a negative penalty.
See Dwyer et al.~\cite{beamsearch} for detailed background and definitions.
 
Note that $m$ cannot be fully `absorbed' when merging with $n$ if either $N(m)\not\subset N(n)$, or $m$ is a leaf of the tree i.e.\ a vertex.
If both cannot be absorbed, then a new module is created that adopts $m$ and $n$ as children and takes away $N(m)\cap N(n)$ as its neighbour set. If only one, say $m$, cannot be absorbed, then $m$ is adopted by $n$, and any shared neighbours are removed from $m$. If both can be absorbed, then one adopts the children of the other and the other is removed, and the operation is truly a full merge. Also note that this is done in the original description by always generating a new parent module, and removing either child if their neighbour sets become empty.

However, measuring the reduction in edges is not the only metric that a greedy algorithm can use to judge the quality of any given merge.
Only using one heuristic often results in many merges with the same score, and so the choice of merge may be arbitrarily chosen from many candidates, some of which may lead directly to local optima. In such situations it can be useful to include additional heuristics to further discriminate between choices.
This is not explicitly mentioned in~\cite{beamsearch}, but is implemented in their provided source code, where they include two other metrics: the total number of modules after the merge, and the number of times power group boundaries are crossed by power edges.

\begin{figure}
    \begin{center}
    \setlength{\tabcolsep}{.15cm}
    \renewcommand{\arraystretch}{1.25}
    \definecolor{grey}{RGB}{170,251,170} 
    
    \begin{tabular}{ |c|c|c|c|c| } 
     \hline
     \multirow{2}{*}{Name ($|E|$)} &
     \multicolumn{2}{c|}{Best $|P|$ $(|G|)$} & \multicolumn{2}{c|}{Worst $|P|$ $(|G|)$} \\ 
     \cline{2-5}
     & only $\kappa_\cap$ & $\kappa_\cap$ and $\kappa_\triangle$ & only $\kappa_\cap$ & $\kappa_\cap$ and $\kappa_\triangle$ \\ 
     \hline\hline
     florentine (20)
        & 11 (4)
        & 11 (4)
        & 11 (6)
        &\cellcolor{grey} 11 (5) \\
     karate (78)
        & 28 (13)\
        & 28 (13)
        & 30 (15)
        &\cellcolor{grey} 29 (13)\\ 
     southern (89)
        & 30 (17)
        &\cellcolor{grey} 27 (21)
        & 37 (16)
        &\cellcolor{grey} 30 (18)\\ 
     dolphins (159)
        & 82 (29)
        &\cellcolor{grey} 81 (30)
        & 87 (28)
        &\cellcolor{grey} 83 (30)\\ 
     lesmis (254)
        & 74 (39)
        &\cellcolor{grey} 72 (41)
        & 79 (39)
        &\cellcolor{grey} 72 (42)\\
     football (613)
        & 282 (83)
        &\cellcolor{grey} 278 (84)
        & 289 (83)
        &\cellcolor{grey} 286 (84)\\
     netsci (914)
        & 355 (187)
        &\cellcolor{grey} 338 (184)
        & 371 (189)
        &\cellcolor{grey} 341 (186)\\
        
     \hline
    \end{tabular}
    \end{center}
    \caption{Some experimental results comparing the quality of resulting power graphs with and without our additional heuristic (Section~\ref{implementation}), taking the best and worst scores over 25 runs.
    $|E|$ is the number of edges in the original graph, compressed into $|P|$ power edges and $|G|$ power groups; shaded boxes indicate the best score between the two methods.
    Adding our new heuristic produces the best or joint best results in all graphs, and also improves the consistency of the output, with a better worst result in all cases.
    Note that the variation between runs in our implementation is due to a pseudorandom order of iteration through
    candidate merges (Fig.~\ref{pgd_pseudo}, line~\ref{pseudo:top}).
    The networks are, from top to bottom, Italian families linked by marriage~\cite{florentine}, members of a karate club~\cite{karate}, women meeting at social events~\cite{southernwomen}, interactions between bottlenose dolphins~\cite{dolphins}, co-occurrence of characters in the musical Les Mis\'erables~\cite{lesmis}, American football games between US colleges~\cite{football}, and coauthorships of scientists working on network theory~\cite{netscience}.
    }
    \label{pgd_results}
\end{figure}

We instead introduce a heuristic that can roughly capture the effects of both, and is simpler to calculate: a penalty for the number of edges that could not be merged. This is defined by
\begin{equation}
    \kappa_\triangle(m, n) = |N(m)\triangle N(n)|
\end{equation}
where $\triangle$ denotes the symmetric difference, i.e.\ any unshared neighbours, between the two sets. This effectively measures the number of edges that cannot ever be merged from that side in future iterations, because only top level modules are considered for merging.
It captures the number of modules because a new parent module is only added if both children have unshared neighbours. It also captures the effect of edges crossing group boundaries, because any unshared edges must cross the boundary of its new parent module after the merge.
Rewarding the first heuristic and punishing the second leaves us with a total score of
\begin{figure}
    \removelatexerror
    \DontPrintSemicolon
    \begin{algorithm}[H]
    \SetKwInOut{Input}{inputs}
    \SetKwInOut{Output}{output}
    \SetKwFor{ForEach}{foreach}{}{}
    \SetKwRepeat{Do}{do}{while}
    \SetKwIF{If}{ElseIf}{Else}{if}{}{else if}{else}{}
    \Input{graph $G=(V,E)$}
    \Output{modules $M= \text{set of pairs } (C,N)$}
    
    $M \leftarrow \{(\emptyset,\,N(v))\ |\ v\in V\}$
    
    \Do{$\kappa_\text{best}>0$}{
        $\kappa_\text{best}\leftarrow0$
        
        \ForEach{\emph{pair of modules} $\{m,n\}\in M_\text{top}\times M_\text{top}$}{
        \label{pseudo:top}
            $\kappa_\text{best} \leftarrow \max(\kappa(m,n),\,\kappa_\text{best})$
            \label{pseudo:kappa}
        }
        $\textsc{merge}(\{m,n\}_\text{best})$
        \label{pseudo:merge}
    }
    
    \caption{\textsc{Greedy power graph decomposition}}
    \label{algorithm:pgd}
    \end{algorithm}
    \caption{Pseudocode for the greedy heuristic power graph construction in Section~\ref{implementation}.
    Each module consists of a set of children $C$, and a set of neighbours $N$.
    The score function $\kappa$ is from Equation~(\ref{kappa}), and the merge operation on line~\ref{pseudo:merge} is described in Section~\ref{implementation}, where any new module is parented to its merged children by adding them to its set of children $C$.
    In practice, we maintain a redundant super-module whose children are the top level modules $M_\text{top}$ on line~\ref{pseudo:top}.}
    \label{pgd_pseudo}
\end{figure} 
\begin{figure}[ht!]
    \removelatexerror
    \DontPrintSemicolon
    \begin{algorithm}[H]
    \SetKwInOut{Input}{inputs}
    \SetKwInOut{Output}{output}
    \SetKwFor{ForEach}{foreach}{}{}
    \SetKwRepeat{Do}{do}{while}
    \SetKwIF{If}{ElseIf}{Else}{if}{}{else if}{else}{}
    \Input{modules $M=\text{set of pairs }(C,N)$}
    \Output{drawing $A$}
        
    $V\leftarrow\emptyset,\,E\leftarrow\emptyset,\,P\leftarrow\emptyset$
    
    \ForEach{\emph{module} $m=(C_m,N_m)\in M$}{
        $V\leftarrow V\cup m$
        
        $E\leftarrow E\cup\{(m,c)\ |\ c\in C_m\}$
        \label{pseudo:directed}
        
        $P\leftarrow P\cup\{(m,n)\ |\ n\in N_m\}$
        \label{pseudo:undirected}
    }
    \ForEach{\emph{vertex} $v\in V$}{
        \If{$|N^+(v)|\geq 2$ \emph{and} $|N^{-*}(v)|\geq 2$}{
        \label{pseudo:inout}
            \textsc{split}($v$)
            \label{pseudo:split}
        }
    }
    
    $A \leftarrow \emptyset$
    
    $\mathbf{X} \leftarrow \textsc{layout}((V,\,E\cup P))$
    \label{pseudo:sgd}
    
    \ForEach{\emph{power edge} $p=\{i,j\}\in P$}{
        
        
        $Q_i\:\leftarrow$ paths $\in (V,E)$ from leaves to $i$
        
        $Q_j\leftarrow$ paths  $\in (V,E)$ from $j$ to leaves
        
        $A\leftarrow A\cup\{\,\textsc{spline}(\mathbf{X}_q)\ |\ q\in Q_i\times Q_j\}$
        \label{pseudo:spline}
    }
    \caption{\textsc{Power-confluent drawing}}
    \label{algorithm:pcd}
    \end{algorithm}
    \caption{Pseudocode for our conversion from a set of power graph modules to a rendered drawing, described in Section~\ref{solution}. Note that the edges on line~\ref{pseudo:directed} are ordered pairs, while those on line~\ref{pseudo:undirected} are unordered. This allows, on line~\ref{pseudo:inout}, for $N^+$ to indicate a set of outgoing neighbours, and $N^{-*}$ a set of incoming neighbours plus any connected by power edges in $P$.
    The split operation on line~\ref{pseudo:split} then moves $N^+$ and $N^{-*}$ to separate routing nodes, as in Figure~\ref{radial}.
    In practice we use a shorter edge length between split routing nodes, and so for line~\ref{pseudo:sgd} use a force-directed layout algorithm that can embed edge lengths~\cite{sgd}.
    The specifics of the spline function on line~\ref{pseudo:spline} are outlined in Section~\ref{splicifics}.}
    \label{pcd_pseudo}
\end{figure}
\begin{equation}
\kappa(m,n) =
w_\cap \kappa_\cap(m,n) -
w_\triangle \kappa_\triangle(m,n)
\label{kappa}
\end{equation}
where $w_\cap$ and $w_\triangle$ determine the relative weight of either heuristic. For example, to construct a modular decomposition, $w_\triangle$ may be set to infinity to forbid any module boundary crossings.
We find that setting $w_\cap=10$ and $w_\triangle=1$ works well in practice; these are the parameters used for the table of results in Fig.~\ref{pgd_results}.
The improvements are small, but are consistently better and with less variance.

It is important to note that this is still a simple greedy heuristic, which does not make any guarantees about the quality of the final result.
The original method in~\cite{beamsearch} gives the option to somewhat alleviate this, by optionally maintaining a priority queue of the best configurations seen so far, along with some dynamic programming to prevent re-evaluating configurations already seen.
A further exploration of this extended algorithm is out of scope for this paper, and Dwyer et al.~\cite{beamsearch} additionally note that the qualitative improvement that results from including this priority queue is minimal.

Our final change involves bringing down the complexity of neighbour set intersection to $O(E)$ using hash sets, which reduces the complexity of the algorithm down slightly to $O(|V|^3|E|)$.
Pseudocode for the above described algorithm can be seen in Fig.~\ref{pgd_pseudo}.
Pseudocode for the subsequent routing graph and B-spline generation process, described in Section~\ref{solution}, can be seen in Fig.~\ref{pcd_pseudo}.

\subsubsection{Directed graphs}
\label{directed}
The definition for a directed confluent drawing is almost the same as the undirected case, except with the caveat that any given path can only `flow' in one direction. This is analogous to preventing trains on tracks from crashing into each other. 
Formally, for a directed confluent drawing $B$, this is extra condition~\cite{dickerson} is:
\begin{itemize}
    \item Locally monotone curves in B may share some overlapping portions, but the edges sharing the same portion of a track must all have the same direction along that portion.
\end{itemize}
This means that directed graphs lead to a slightly different power-to-routing graph conversion, where the description in~\cite{bach} says \emph{``The only difference is that we create---as necessary---two junctions for each group, one for incoming and one for outgoing edges."}
Here we will elaborate on this short description, as simply adding another routing node when its corresponding group has both incoming and outgoing edges is not enough to guarantee that all portions will only flow in one direction.
This is because any power edges at a power group are propagated down the hierarchy, so that even if flows are correctly directed into separate junctions at one group, they may still clash further down the tree.
Therefore, all descendants of any such group must also have two routing nodes: one for outgoing edges (flowing up the hierarchy) and one for incoming edges (flowing down the hierarchy).

Our description of the power graph construction algorithm is also easily applied to directed graphs, by splitting neighbour sets into incoming and outgoing edges. This is how the original algorithm is described by Dwyer et al.~\cite{beamsearch}.
In other words, the formulations for the heuristics instead become
\begin{align}
    &\kappa_\cap(m,n) = |N^+(m)\cap N^+(n)| + |N^-(m)\cap N^-(n)|,\\
    &\kappa_\triangle(m,n) = |N^+(m)\triangle N^+(n)| + |N^-(m)\triangle N^-(n)|
\end{align}
where $N^+$ is a set of outgoing edges, and $N^-$ a set of incoming edges.

\section{Discussion}
\label{classification}
Here we will discuss the third condition in the confluent definition, which requires planarity. Unfortunately, the original method does not offer any guarantee of planarity due to the use of a force-directed method to lay out the graph.
While the authors do recognize this, and make the distinction that their drawings are `non-planar confluent', the loss of this condition means that almost any drawing, with or without curved edges, satisfies a now-trivial definition.
There also exist graphs that have been proven to not admit any confluent drawing~\cite{dickerson}, and so it is impossible for any algorithm to produce confluent drawings for general graphs.
However, the planarity condition can be relaxed in the context of finding a drawing that reduces the number of crossings, for example in layered confluent drawings \cite{layered}. Bach et al.~\cite{bach} do not explicitly address this question, but in practice the approach can greatly reduce the number of crossings, making it a practical method to enhance readability.

Furthermore, it is easy to see that a planar routing graph can always lead to a planar drawing, as long as we are careful to avoid crossings at routing nodes (which is always possible because every routing node is simply a single junction). If we assume that there are no redundant routing edges without adjacencies routed through them, this means that the complete confluent definition can be satisfied if and only if the routing graph is also planar.
This naturally implies a new subset of confluent drawings, much like the $\Delta$-confluent~\cite{delta} or strict confluent~\cite{strict} subclasses, which can be found only through a planar power-to-routing graph conversion. We suggest naming such drawings as \emph{power-confluent}.

\begin{figure}
    \centering
    \includegraphics[width=\linewidth]{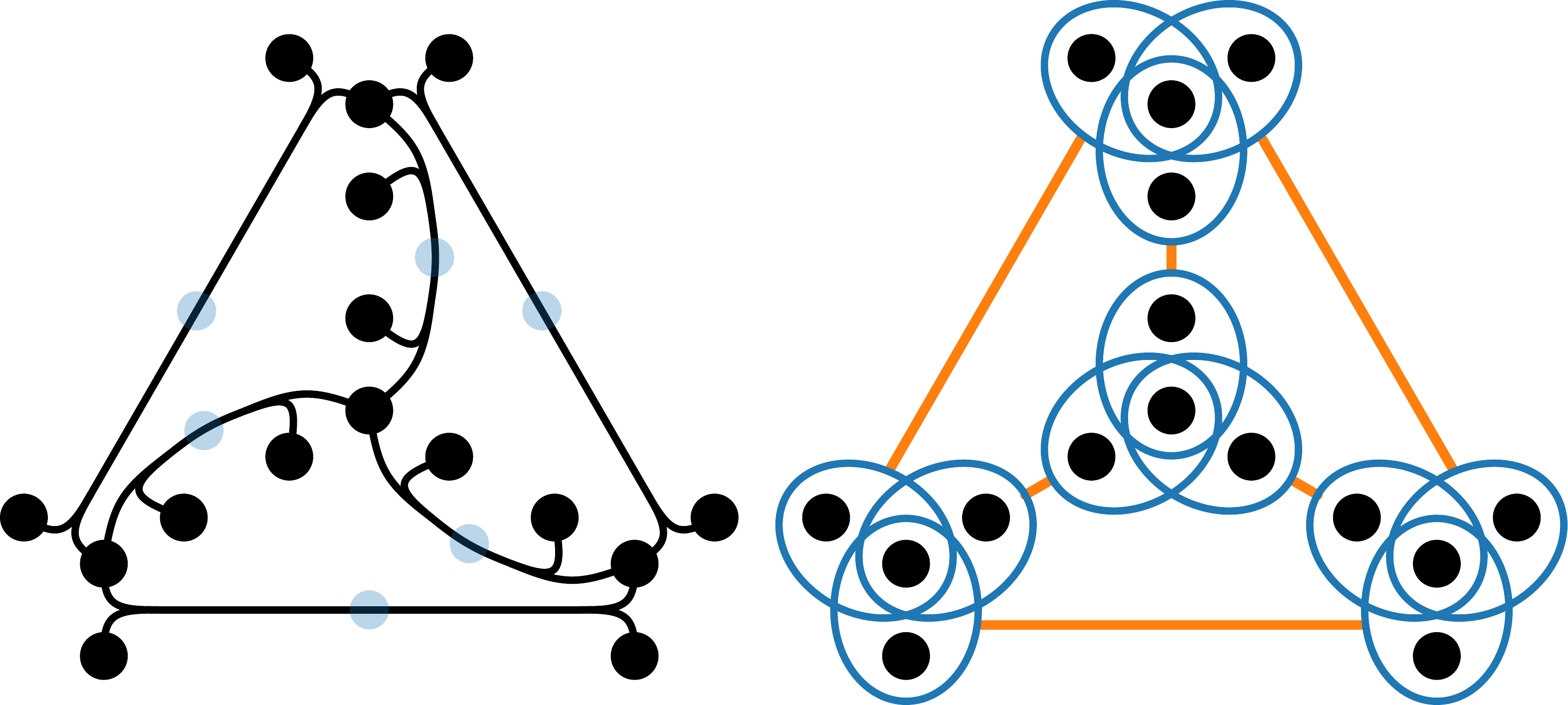}
    \caption{An example of a strict confluent drawing that cannot be reverse-engineered into a valid power graph, unless power group boundaries are allowed to overlap. On the left is the confluent drawing (with junctions marked by transparent blue circles), and on the right is a power graph that shows all possible groupings.
    Since there are six junctions, at least six power groups are required in the corresponding power graph; however, there is a maximum of four that will not overlap, as there can only be one for each `corner' of the `tetrahedron'. This is due to the middle vertex of each `corner' sharing junctions with all three surrounding it, which means that grouping any of the three valid pairings will block out the other two.
    Note that the graph itself is not necessarily impossible to draw in a power-confluent manner, but that this particular drawing could never result from the algorithm presented in this paper.}
    \label{strict}
\end{figure}

The definition of strict confluent drawing has only the additional restriction that there can be at most one smooth path between vertices, and there cannot be any paths from a vertex to itself~\cite{strict}. 
For the power-confluent case, because the mapping of edges to power edges is surjective~\cite{royer}, each edge can only correspond to a single path through the power graph hierarchy, and therefore also through the routing graph behind the confluent drawing. Every power-confluent drawing without self-loops is therefore also strict confluent.
The reverse is not true, because another condition of power graphs is that groups must be disjoint~\cite{royer}, i.e.\ their boundaries may not overlap. There is a family of confluent drawings that cannot result from power graphs unless this condition is broken, for example the tetrahedron-like structure in Fig.~\ref{strict}.
Power-confluent is therefore a strictly stronger condition than strict confluent.

\subsection{Future Work}
Further directions could involve developing methods to find power-confluent drawings, by guiding the search algorithm towards solutions that produce planar routing graphs.
Methods such as Monte Carlo or A* search may prove useful for either finding such drawings, or just improving the greedy search presented here.

On the theoretical side, it may be the case that relaxing the non-overlapping group boundary condition, as explored by Ahnert~\cite{generalised}, could result in an equivalent classification to strict confluent drawing.
We leave the potential proof or refutation of this equivalence as an open question, along with determining the exact relationship between power decomposition and confluent drawing in general.

To finish on a practical note, a more tailored layout algorithm than standard force-directed methods will be necessary for the algorithm to become a truly practical tool, as we find that layouts can often become tangled and unreadable. The layout function (Fig.~\ref{pcd_pseudo}, line~\ref{pseudo:sgd}) is currently only given the routing graph as input, but may benefit from extra information, such as which edges are power edges and which are hierarchical. An effective radial layout that can avoid crossings, like the one shown in Fig.~\ref{radial} but automatic, may also offer a superior solution.

\section*{Acknowledgements}
We would like to thank the anonymous reviewers whose comments helped to improved this paper.

\setcounter{figure}{0}
\renewcommand{\thefigure}{A\arabic{figure}}
\begin{figure}
    \centering
    \includegraphics[width=\linewidth]{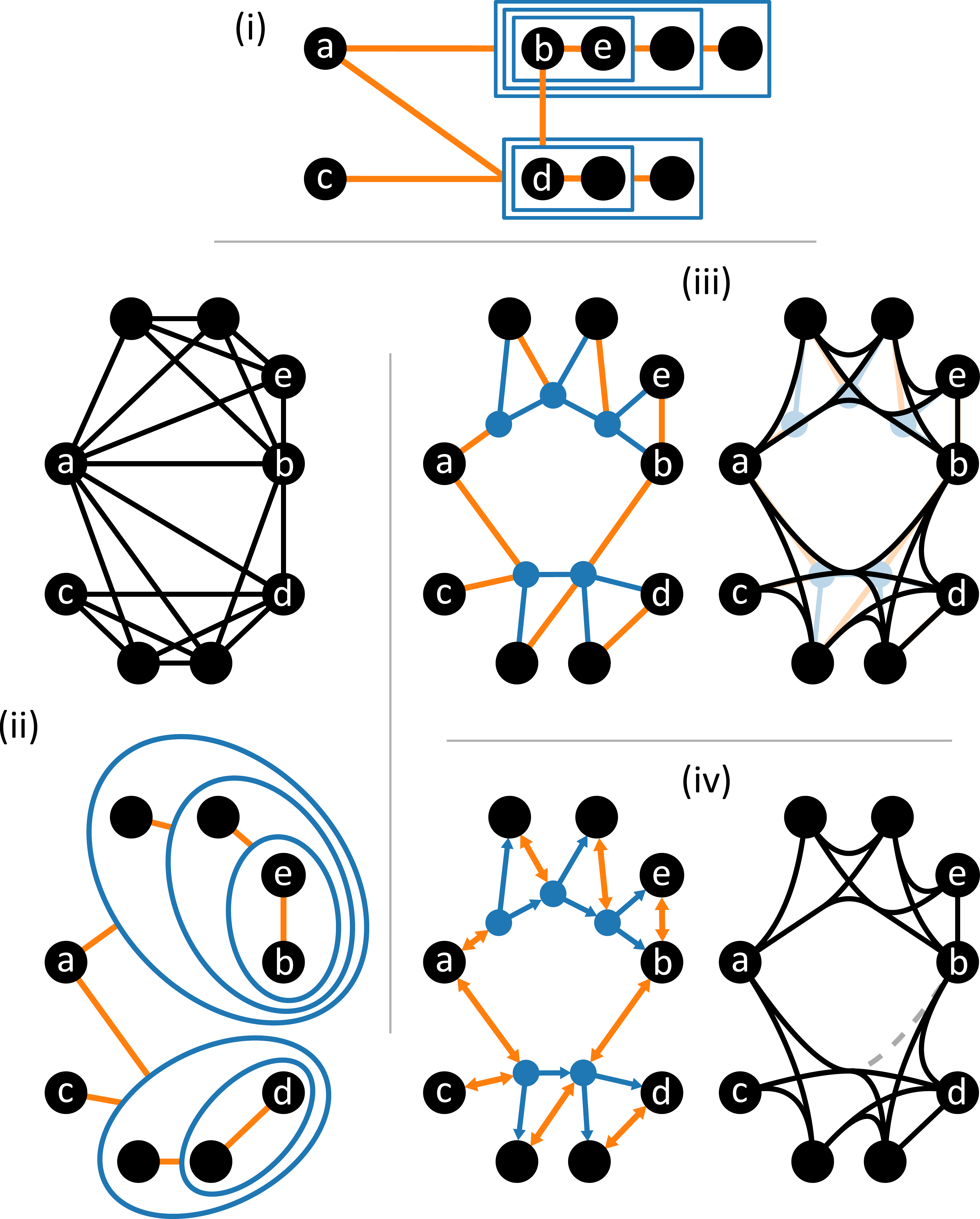}
    \caption{A full example of Fig.~\ref{overlap}, bottom, without any redundant power groups.
    (i): an example of the power graph in the same layout as Fig.~\ref{overlap}, where the nested structure of groups comes from a clique structure.
    (ii) top: a conventional node-link layout of the graph, bottom: the same power graph as in (i), but using the same layout as above. 
    (iii) left: the resulting routing graph from the method of Bach et al.~\cite{bach}, right: the resulting drawing when splines take their shortest paths through this routing graph. This results in the edge $\{a,b\}$ being routing downwards through the wrong direction, which causes the edge $\{c,b\}$ to falsely appear to exist.
    The edge $\{a,e\}$ also has two equal-length shortest paths, either through the line shown or downwards through $b$, albeit this specific case can easily be prevented by only preventing original nodes from being used as intermediate control points, as the original authors appear to have done in~\cite[Fig.~2(c)]{bach}  for the edge $\{u,w\}$.
    (iv) left: the result of using our new method (Section~\ref{solution}) of retaining the hierarchical structure of power groups through directed edges, right: the resulting drawing where $\{a,b\}$ is routed through the three correct upward routing nodes. The previously incorrect routing is marked by a dashed line.
    }
    \label{overlap_sm}
\end{figure}

\end{document}